\documentclass{appolb}
\usepackage{epsfig}
\def\Journal#1#2#3#4{{#1} {\bf #2}, #3 (#4)}

\def\NPB{{\em Nucl. Phys.} B}
\def\PLB{{\em Phys. Lett.}  B}
\def\PRL{\em Phys. Rev. Lett.}

\def\EPJC{{\em Eur. Phys. J.} C}

\newcommand{\pom}{I\!\!P}

\begin{document}
\title{Diffractive interactions in ep collisions%
\thanks{Presented at Photon2005, Warsaw}%
}
\author{Isabell-Alissandra Melzer-Pellmann
\address{DESY, ZEUS group, Notkestr. 85, 22607 Hamburg}
}
\maketitle
\begin{abstract}
The H1 and ZEUS experiments are measuring diffractive interactions
in $ep$ collisions at HERA. Performing QCD fits of these data with 
NLO DGLAP, diffractive parton distribution functions can be calculated. 
These diffractive PDFs can be used to test QCD factorisation with 
dijet and charm data.
\end{abstract}
  
\section{Introduction}
Quantum chromodynamics (QCD) describes the strong interactions between 
quarks and gluons at small distances very well, where the strong coupling 
constant $\alpha_S$ is small and the calculation becomes perturbative.
However, the calculation of total cross sections, usually dominated by 
long range forces or "soft interactions", needs more understanding. A 
fraction of these interactions are characterized by the exchange of a 
color singlet with vacuum quantum numbers. These "diffractive" 
interactions are well described by Regge theory, where a leading 
("Pomeron") trajectory with vacuum quantum numbers is exchanged in the
$t$-channel.
\begin{figure} [t]
  \begin{center}
    \epsfig{figure=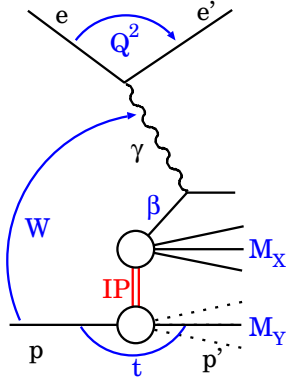,width=0.3\textwidth}
  \end{center}
  \caption{Diffractive $ep$ scattering. 
    \label{Melzer:fig1}}
\end{figure}
Figure~\ref{Melzer:fig1} shows a sketch of the generic diffractive process in $ep$ scattering
at HERA, displaying the most important variables. The hard scale of the 
interaction is given by the photon virtuality $Q^2$, $W$ is the $\gamma p$ 
centre of mass energy, and $t$ the squared four-momentum transfer at the proton 
vertex, where the proton has the four-momentum $p$. The colorless diffractive 
object ($\pom$) carries the momentum fraction $x_{\pom}$, and the quark struck 
by the photon with four-momentum $q$ has the fraction $\beta$ of the
momentum of the diffractive exchange. They are defined as follows:
\begin{equation}
x_{\pom} = \frac{q \cdot (p-p')}{q \cdot p} \approx \frac{Q^2 + M_X^2}{Q^2+W^2} \:, \: \:
\beta = \frac{Q^2}{2 q \cdot (p-p')} \approx \frac{Q^2}{Q^2 + M_X^2} = \frac{x}{x_{\pom}} \:,
\end{equation}
with $x$ being the Bjorken-$x$. The diffractively produced system $X$ has 
the mass $M_X$. The proton can stay intact and get scattered with the four-momentum
$p'$ or dissociate in a system $Y$ with mass $M_Y$.
\section{Event selection}
There are different methods to select diffractive events. A very clean way is
to detect the scattered proton in the forward direction. H1 and ZEUS have 
proton spectrometers placed along the beam line in the direction of the proton 
beam. In addition to being free
from proton dissociation background, the advantage of this selection is the 
possibility to measure the momentum transfer $t$. On the other hand, this method 
is limited by statistics due to small acceptance.

A high statistics sample can be obtained taking advantage of the characteristic 
properties of the final state, where we observe a large gap in rapidity between 
the leading proton (or in case of dissociation the leading baryonic system) and 
the photon dissociation system due to the colorless exchange. This gap can
be identified by the abscence of activity in the forward part of the calorimeter.
The residual proton dissociation background is about 9\,\% for masses
$M_Y < 1.6$ GeV and can be subtracted statistically. 

The third method to extract diffractive is using the fact that the $M_X$ 
distributions behave differently for diffractive and non-diffractive data. 
Monte Carlo simulations show that the diffractive contribution is almost flat in
$\ln M_X^2$ while the non-diffractive contribution is exponentially falling 
for decreasing $M_X$. The diffractive data is extracted with a fit of the 
variable $\ln M_X^2$. The proton dissociation background gets subtracted for
masses of $M_Y > 2.3$ GeV \cite{ZEUS05}.

\section{Cross sections and extraction of PDFs}
The differential diffractive cross section $\sigma^D$ can be defined 
\begin{equation}
\frac{d^3 \sigma^D(x_{\pom},x,Q^2)}{d x_{\pom}\,dx\,dQ^2} = \frac{4 \pi \alpha^2}{x Q^4} 
\left((1-y+\frac{y^2}{2}\right) \sigma_r^{D(3)} (x_{\pom},x,Q^2)\, ,
\end{equation}
with the reduced cross section $\sigma_r^{D}$, which is related to the 
diffractive structure functions $F_2^{D}$ and $F_L^{D}$, neglecting 
contributions from the $Z^0$ exchange, by

\begin{equation}
\sigma_r^{D(3)} = F_2^{D(3)} - \frac{y^2}{1+(1-y)^2} F_L{D(3)}     \, .
\end{equation}

\begin{figure} [b]
  \begin{minipage}[b]{0.48\linewidth} 
    \centering
    \epsfig{figure=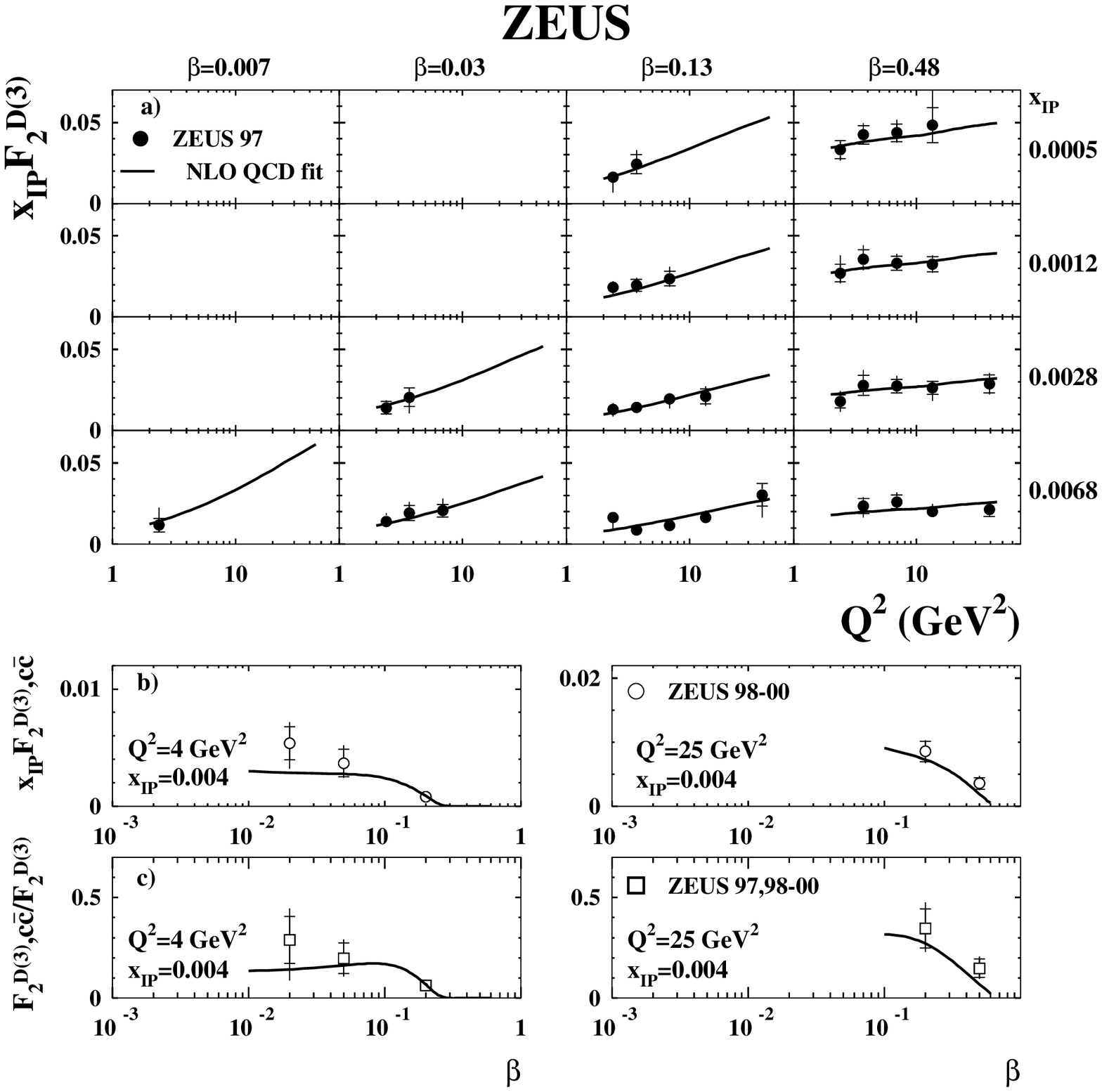,width=.93\textwidth}
    \caption{NLO QCD fits from ZEUS.\label{Melzer:fig3}}
  \end{minipage}
  \hspace{0.2cm} 
  \begin{minipage}[b]{0.48\linewidth}
    \centering
    \label{Melzer:fig4}
    \epsfig{figure=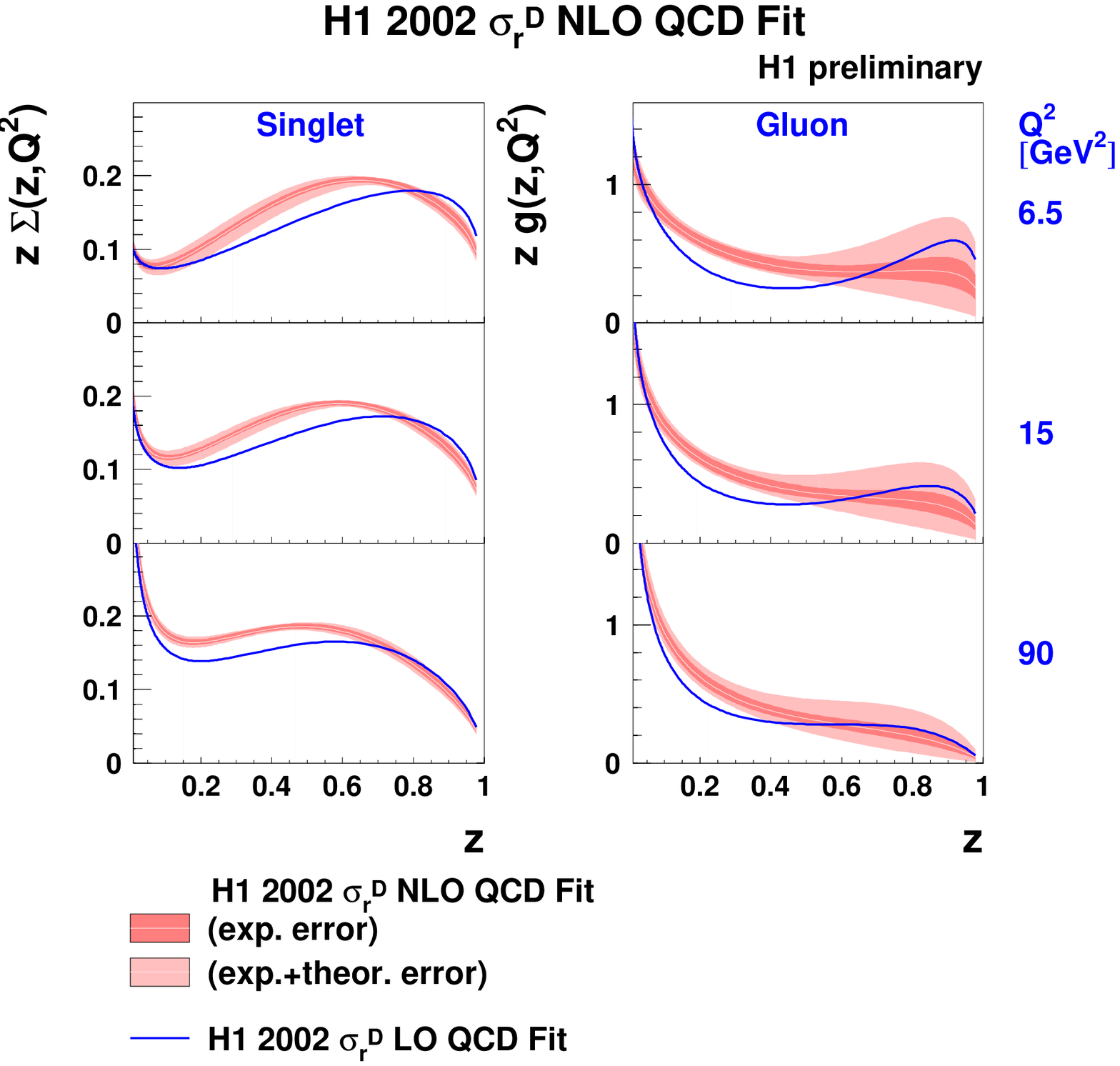,width=.93\textwidth}
    \caption{Diffractive PDFs from H1.}
  \end{minipage}
\end{figure}

Assuming Regge factorisation, QCD fits have been performed by H1 \cite{H102} and 
ZEUS \cite{ZEUS04}, using the DGLAP formalism to evolve the non-perturbative 
diffractive parton densities (dPDFs). The diffractive exchange is modeled in terms
of a light flavor singlet and a gluon distribution, parameterized by a set of
polynomials at a starting scale $Q_0^2 = 3$ GeV$^2$ in case of H1. ZEUS includes
also diffractive charm data in the fit, starting at $Q_0^2 = 2$ GeV$^2$. H1 treats 
the charm quark in the massive scheme via boson-gluon fusion processes with 
$m_c = 1.5 \pm 0.1$ GeV, while ZEUS uses the Thorne-Roberts variable flavor number
(TRVFN) scheme with $m_c = 1.45$ GeV. Both experiments find a large momentum fraction
exchanged by gluons of $\sim 75 \pm 15$ \% (H1) and 
$82 \pm 8 {\rm{(stat)}} ^{+5}_{-16} {\rm{(syst)}}$ \% (ZEUS) at the initial scale. 
Figure~\ref{Melzer:fig3} shows the result of the fit for ZEUS and Fig.~\ref{Melzer:fig4} 
shows the diffractive PDFs from H1.

Using these PDFs, one can perform tests of the validity of the assumptions made for 
the QCD analysis, primarily the QCD factorisation. 

\section{Dijet and charm analyses using the diffractive PDFs}
Since the PDFs are gluon dominated, there is a special interest in processes which
are sensitive to photon-gluon interactions, like dijet and heavy flavor
processes. The longitudinal fraction of the Pomeron momentum carried by the emitted
gluon $z_{\pom}$ is determined using the invariant mass $M_{12}$ of the $q\bar{q}$ 
system.

\subsection{Diffractive dijets and charm in DIS}
\begin{figure} [b]
  \begin{minipage}[b]{0.48\linewidth} 
    \centering
    \epsfig{figure=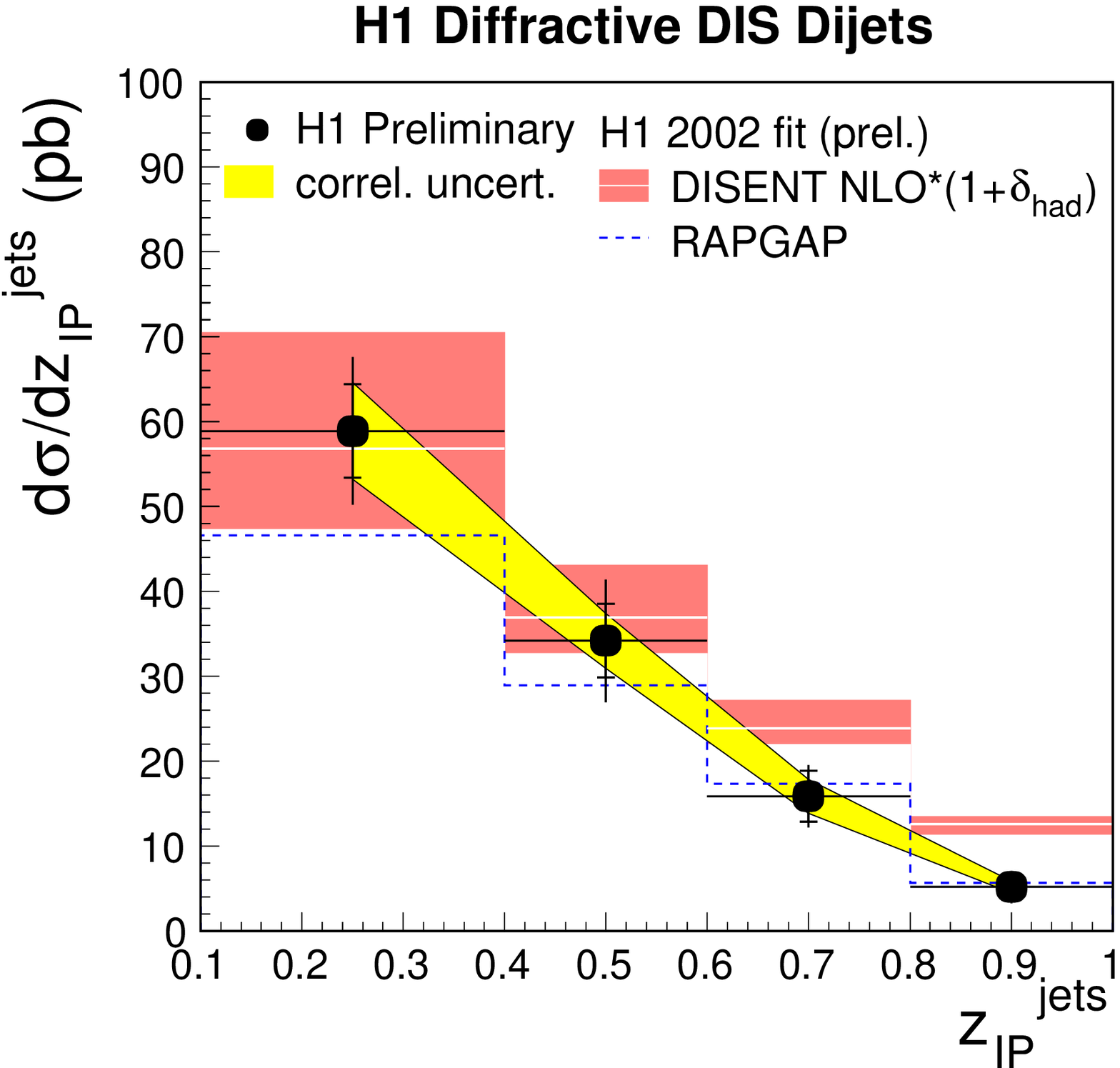,width=1.\textwidth}
    \caption{Diffractive dijets in DIS.
      \label{Melzer:fig5}}
  \end{minipage}
  \hspace{0.2cm} 
  \begin{minipage}[b]{0.48\linewidth}
    \centering
    \epsfig{figure=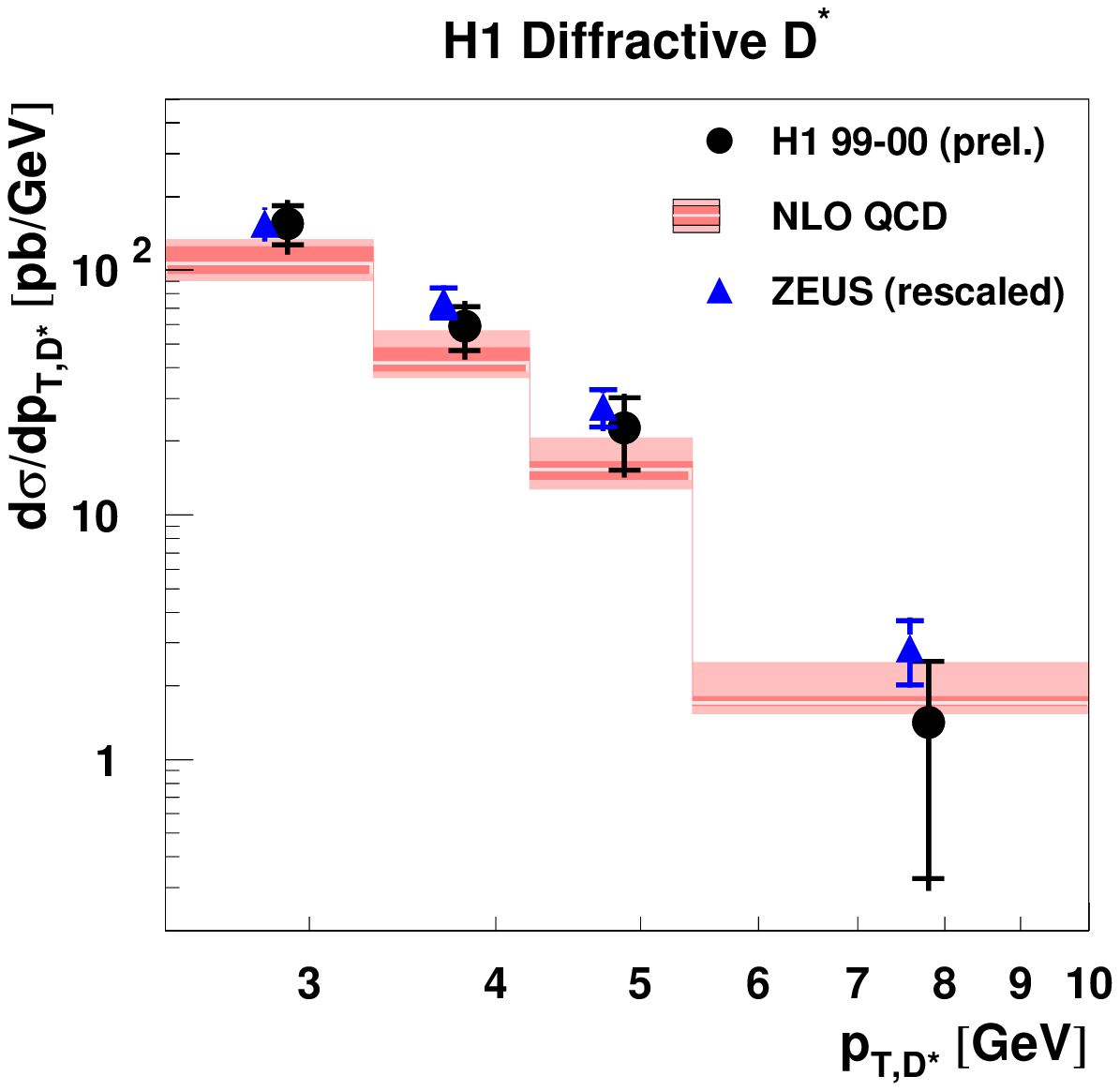,width=1.\textwidth}
    \caption{Diffractive $D^*$ in DIS.
      \label{Melzer:fig6}}
  \end{minipage}
\end{figure}

Figure~\ref{Melzer:fig5} shows the measured dijet cross section as a function of 
$z_{\pom}$ from H1 \cite{H104}, compared to the {\sc Disent} NLO calculation 
\cite{DIS97} interfaced to the H1 PDFs shown above. The error band of the NLO 
calculation is estimated varying the renormalisation scale by factors 0.5 and 2. 
Also shown is  the {\sc Rapgap} \cite{RAP95} prediction, which contains parton 
showers and is based on the LO PDFs from the same fit. ZEUS has done a similar 
measurement \cite{ZEUS05b}, comparing the data to NLO QCD calculations with the 
H1 PDFs and the ZEUS PDFs from the measurements shown above. These describe the 
data well, while a third NLO calculation using the GLP fit \cite{GLP05} underestimates
the data.

Both ZEUS \cite{ZEUS03} and H1 \cite{H104b} have measured diffractive $D^*$ production
in DIS.
The cross section as a function of the transverse momentum $p_{T,D^*}$ of the $D^*$
is shown in Fig.~\ref{Melzer:fig6}, compared to an NLO calculation using a 
diffractive version of the program {\sc Hvqdis} \cite{Har95}. The inner error band
represents the renormalization scale uncertainty varied by factors 1/4 to 4, while 
the outer error band shows the total uncertainty, including variations of the charm 
mass from 1.35 to 1.65 GeV and of the parameter of the Peterson fragmentation 
function $\epsilon$ from 0.035 to 0.100, added in quadrature.

Both, the dijet as well as the $D^*$ production in DIS are reasonably well 
described by the NLO calculations using the H1 PDFs, supporting the assumption 
that factorisation holds for diffractive reactions in DIS. 

\subsection{Diffractive dijets and $D^*$ in photoproduction}
\begin{figure} [b]
  \begin{center}
    \epsfig{figure=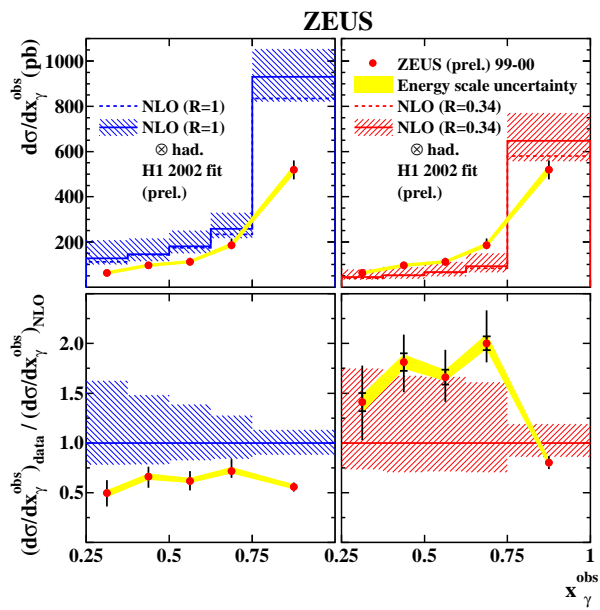,width=0.7\textwidth}
  \end{center}
  \caption{Diffractive dijets in photoproduction. 
    \label{Melzer:fig7}}
\end{figure}
 
Applying a similar QCD calculation to diffractive dijet production at Tevatron,
the observed rate is overestimated by a factor of 3 to 10, depending on the chosen
diffractive PDF \cite{CDF00}. This factorisation breaking is explained by
Kaidalov et al. \cite{Kai03} as being caused by secondary interactions of additional
spectator quarks in the proton remnant, which are not present in virtual photons
(in DIS). In photoproduction though, the photon can either participate directly
in the hard scattering subprocess ("direct photon"), or fluctuate into partons
("resolved photon"). In this case only a part $x_{\gamma} < 1$ of the photon momentum
enters the hard scattering. This resolved photon part is similar to hadron-hadron 
scattering and should therefore be suppressed, as seen at Tevatron. Kaidalov et al.
\cite{Kai03} have predicted a suppression factor of 0.34 for the resolved photon 
contribution in diffractive dijet photoproduction. 

H1 \cite{H104} and ZEUS \cite{ZEUS04b} have measured diffractive dijets in
photoproduction and compared the result to NLO QCD calculations. Figure~\ref{Melzer:fig7} 
shows the cross section and the ratio of the cross section over the NLO
prediction as a function of $x_{\gamma}^{\rm obs}$. Although the shape is described by 
the NLO calculation quite well, the cross section is overestimated in all bins. Scaling 
the resolved part by the factor 0.34 does not describe the shape. This suggests that a 
global suppression is more likely than a resolved photon suppression only.

\begin{figure} [t]
  \begin{center}
    \begin{tabular}{cc}
    \mbox{\epsfig{figure=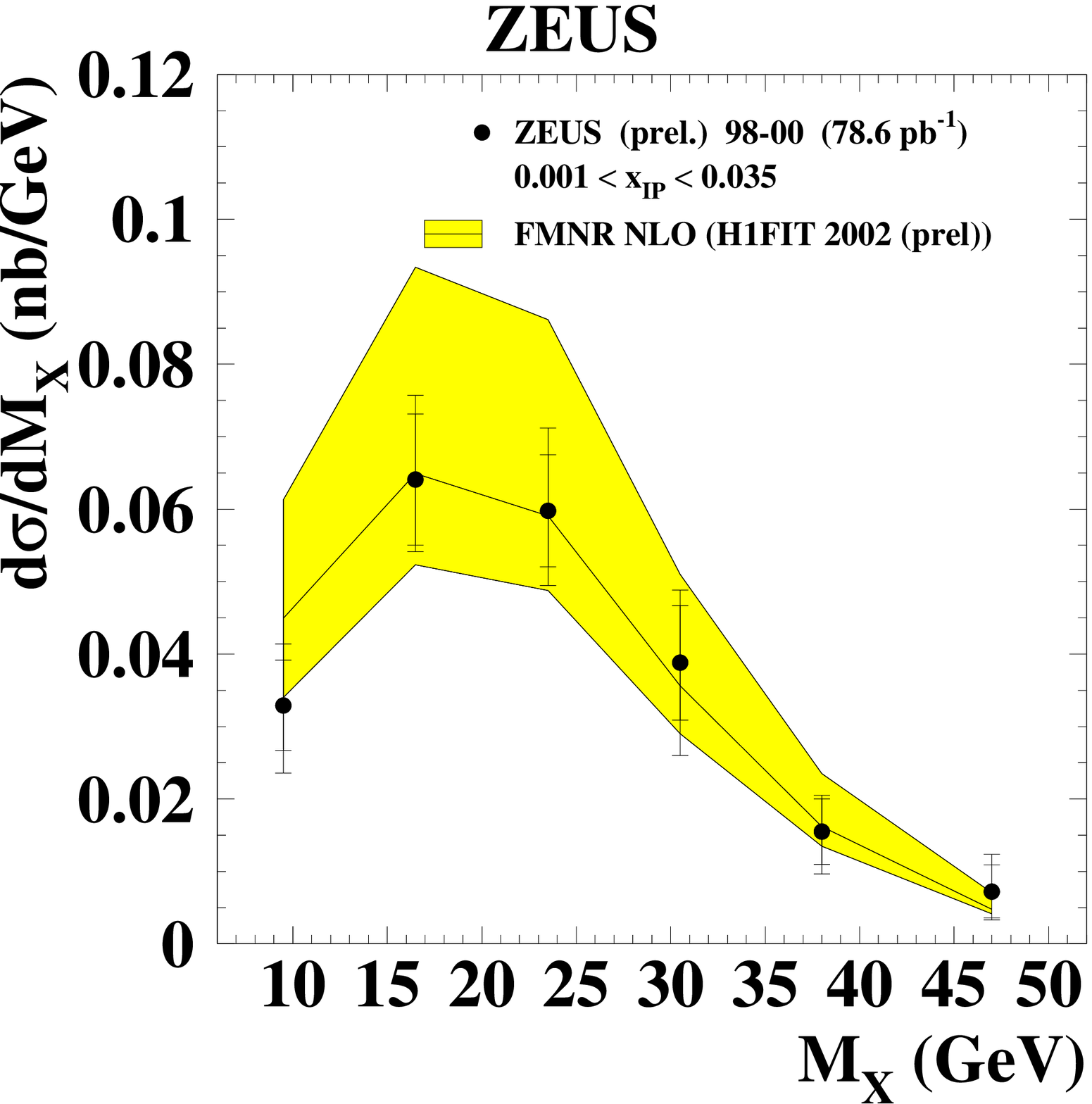,width=0.45\textwidth}} &
    \mbox{\epsfig{figure=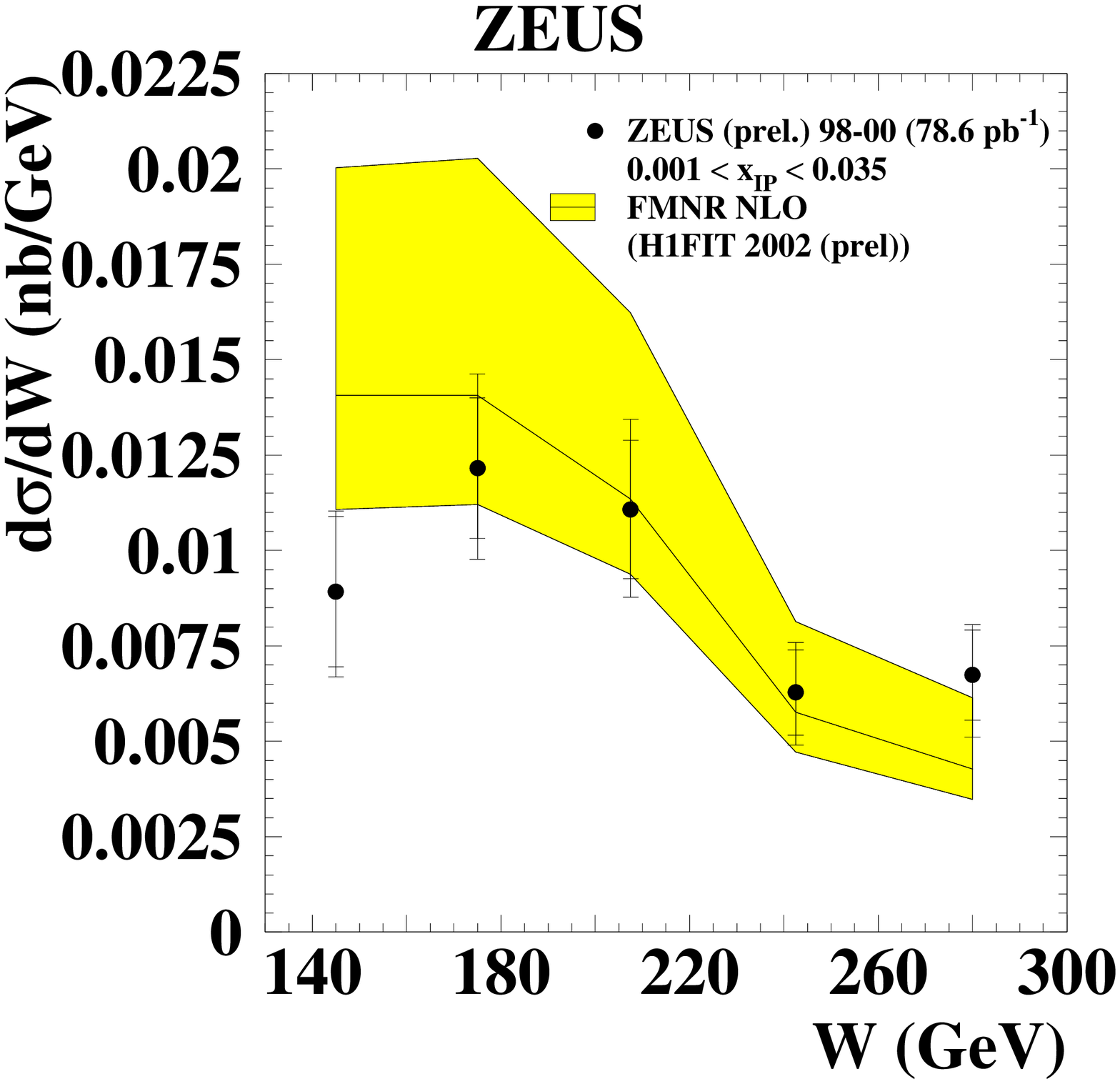,width=0.45\textwidth}} 
    \end{tabular}
  \end{center}
  \caption{Diffractive $D^*$ in photoproduction. 
    \label{Melzer:fig8}}
\end{figure}

ZEUS has recently measured diffractive $D^*$ in photoproduction \cite{ZEUS05c}. 
Figure~\ref{Melzer:fig8} shows the cross section in bins of $M_X$ and $W$ in comparison to
NLO calculations using the FMNR \cite{Frix94} program with the H1 PDFs shown above. 

The error band includes, in addition to the variation of the charm mass 
($m_c = 1.5 \pm 0.2$ GeV), variations of the fragmentation and renormalisation scale 
by factors 0.5 and 2. The data are both in shape and in the total normalisation well 
described.

This is not necessarily a contradiction to the dijet results, taking into account that
the inclusive $D^*$ in photoproduction cross section is underestimated in the NLO 
calculations \cite{ZEUS02} by approximately the same amount the diffractive dijet
cross section is overestimated. The NLO calculations for the inclusive dijets in 
photoproduction \cite{ZEUS03b} describe the data well in shape and magnitude.

\section{Conclusions}
Diffractive exchange contributes a substantial part to the deep-inelastic $ep$ 
scattering at HERA. Assuming Regge factorisation, the nature of the diffractive
exchange can be studied with QCD fits to inclusive diffractive data. These fits
show that diffractive exchange is dominated by gluons, contributing about 75 \% of 
the exchanged momentum.

Using the PDFs obtained from the H1 and ZEUS QCD fits shown above, the 
factorisation assumption is successfully tested for the diffractive production 
of dijets and heavy flavor in DIS. Using the GLP PDFs, the data are 
underestimated, leading to the conclusion that the PDFs probably have a large 
uncertainty. Applying a similar QCD calculation to 
diffractive dijet production at Tevatron, the observed rate is overestimated, 
which is explained by secondary interactions of spectator partons in hadron-hadron
interactions. This suppression should also be visible in photoproduction at HERA,
in part of the reactions where the photon is resolved, but there is no clear
picture yet. The diffractive dijets in photoproduction not only a suppression
of the resolved part, but an overall suppression of the order of 0.5. The
diffractive $D^*$ data seem to be described by the NLO calculations, but as the
inclusive $D^*$ in photoproduction is underestimated by similar NLO calculations,
one has to be careful with the interpretation of this result.

\end{document}